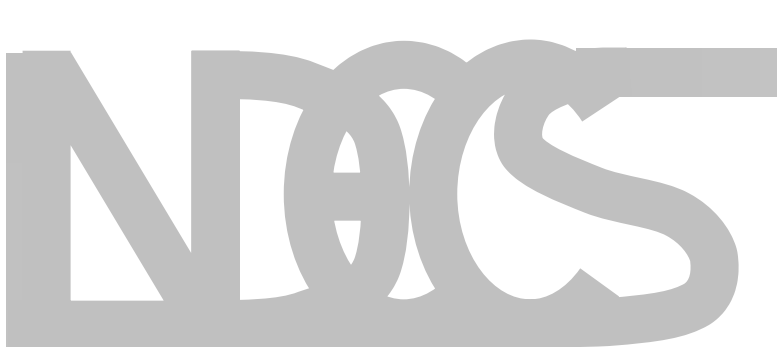

# UNIVERSAL SCIENCE OF COMPLEXITY: CONSISTENT UNDERSTANDING OF ECOLOGICAL, LIVING AND INTELLIGENT SYSTEM DYNAMICS

**Andrei Kirilyuk***

Solid State Theory Department, Institute of Metal Physics
Kiev, Ukraine

## SUMMARY

A major challenge of interdisciplinary description of complex system behaviour is whether real systems of higher complexity levels can be understood with at least the same degree of objective, "scientific" rigour and universality as "simple" systems of classical, Newtonian science paradigm. The problem is reduced to that of arbitrary, many-body interaction (unsolved in standard theory). Here we review its causally complete solution, the ensuing concept of complexity and applications. The discovered key properties of dynamic multivaluedness and entanglement give rise to a qualitatively new kind of mathematical structure providing an exact version of real system behaviour. The extended mathematics of complexity contains a truly universal definition of dynamic complexity, randomness (chaoticity), classification of all possible dynamic regimes, and a unifying principle of any system dynamics and evolution, the universal symmetry of complexity. Every real system has a non-zero (and actually high) value of unreduced dynamic complexity determining, in particular, "mysterious" behaviour of quantum systems and relativistic effects causally explained now as unified manifestations of complex interaction dynamics. The observed differences between various systems are due to different regimes and levels of their unreduced dynamic complexity. We outline applications of universal concept of dynamic complexity emphasising the case of "truly complex" systems from higher complexity levels (ecological and living systems, brain operation, intelligence and consciousness, autonomic information and communication systems) and show that the urgently needed progress in social and intellectual structure of civilisation inevitably involves qualitative transition to unreduced complexity understanding (we call it "revolution of complexity"). It realises a new kind of knowledge combining intrinsic interdisciplinarity, universality, causal completeness, and possibility of rigorous expression in various, universally accessible forms.

## KEY WORDS

## CLASSIFICATION

*Corresponding author, $\eta$: Andrei.Kirilyuk@Gmail.com;
Institute of Metal Physics, 36 Vernadsky Avenue, 03142 Kiev-142, Ukraine

## INTRODUCTION

A major challenge of emerging interdisciplinary studies of complex system behaviour is whether any higher-level system dynamics can be described at least with the same degree of rigour and universality as that of usual, Newtonian science with respect to lower-level, "mechanical" system dynamics. The problem is actually reduced to that of real, many-body interaction dynamics, which remains unsolved (or "nonintegrable") already starting from three interacting bodies (for arbitrary interaction potential). Here we review a recently proposed, universally nonperturbative solution of arbitrary interaction problem, the ensuing universal, realistic concepts of dynamic complexity, chaoticity, self-organisation, fractality and their applications to real systems, especially those of higher complexity levels (living, intelligent, social and ecological systems) [1-15]. We reveal the universal symmetry of complexity [1-5] as a unifying principle of any system dynamics and evolution (and thus extension of all particular laws), as well as intrinsic interdisciplinarity of universal science of complexity confirmed by various problem-solving applications, from fundamental physics to the humanities [1].

## UNREDUCED INTERACTION DYNAMICS AND UNIVERSAL COMPLEXITY DEFINITION

Consider arbitrary, many-body interaction described by a general enough dynamic equation called here *existence equation* and simply reflecting initial system configuration [1-10]:

$$\left\{ \sum_{\mathrm{k}=0}^{N} \left[ h_{\mathrm{k}}\left(q_{\mathrm{k}}\right) + \sum_{\mathrm{l}>\mathrm{k}}^{N} V_{\mathrm{kl}}\left(q_{\mathrm{k}}, q_{\mathrm{l}}\right) \right] \right\} \Psi\left(Q\right) = E \Psi\left(Q\right) \ , \tag{1}$$

where $h_{\mathrm{k}}(q_{\mathrm{k}})$ is the generalised Hamiltonian of the k-th system component (degrees of freedom $q_{\mathrm{k}}$), $V_{\mathrm{kl}}(q_{\mathrm{k}}, q_{\mathrm{l}})$ is the potential of interaction between the k-th and l-th components, $\Psi(Q)$ is the system state-function depending on all degrees of freedom, $Q \equiv \{q_0, q_1, ..., q_N\}$, $E$ is the generalised Hamiltonian eigenvalue, and summations are performed over all ($N$) system components. The Hamiltonian equation form does not involve any real limitation and can be rigorously derived as universal expression of real system dynamics [1-10], where generalised Hamiltonians express suitable measures of complexity (see below). One can rewrite eq. (1) in another form, where a degree of freedom, $q_0 \equiv \xi$, is separated as it represents common system component or measure (such as position of other, localised components):

$$\left\{ h_0\left(\xi\right) + \sum_{\mathrm{k}=1}^{N} \left[ h_{\mathrm{k}}\left(q_{\mathrm{k}}\right) + V_{0\mathrm{k}}\left(\xi, q_{\mathrm{k}}\right) + \sum_{\mathrm{l}>\mathrm{k}}^{N} V_{\mathrm{kl}}\left(q_{\mathrm{k}}, q_{\mathrm{l}}\right) \right] \right\} \Psi\left(\xi, Q\right) = E \Psi\left(\xi, Q\right) \ , \tag{2}$$

where $Q \equiv \{q_1, ..., q_N\}$ and $k, l \geq 1$ (also everywhere below).

It is convenient to express the problem in terms of free component eigenfunctions $\{\varphi_{\mathrm{kn_k}}(q_{\mathrm{k}})\}$ and eigenvalues $\{\varepsilon_{\mathrm{n_k}}\}$:

$$h_{\mathrm{k}}\left(q_{\mathrm{k}}\right) \varphi_{\mathrm{kn_k}}\left(q_{\mathrm{k}}\right) = \varepsilon_{\mathrm{n_k}} \varphi_{\mathrm{kn_k}}\left(q_{\mathrm{k}}\right) \ , \tag{3}$$

$$\Psi\left(\xi, Q\right) = \sum_{\mathrm{n} \equiv \left(\mathrm{n}_1, \mathrm{n}_2, ..., \mathrm{n}_N\right)} \psi_{\mathrm{n}}\left(q_0\right) \varphi_{1\mathrm{n}_1}\left(q_1\right) \varphi_{2\mathrm{n}_2}\left(q_2\right) ... \varphi_{N\mathrm{n}_N}\left(q_N\right) \equiv \sum_{\mathrm{n}} \psi_{\mathrm{n}}\left(\xi\right) \Phi_{\mathrm{n}}\left(Q\right) \ , \tag{4}$$

where $\Phi_{\mathrm{n}}\left(Q\right) \equiv \varphi_{1\mathrm{n}_1}\left(q_1\right) \varphi_{2\mathrm{n}_2}\left(q_2\right) ... \varphi_{N\mathrm{n}_N}\left(q_N\right)$ and $\mathrm{n} \equiv \left(\mathrm{n}_1, \mathrm{n}_2, ..., \mathrm{n}_N\right)$ runs through all eigenstate combinations. Inserting eq. (4) into eq. (2) and performing usual eigenfunction separation (e.g. by taking scalar product), we obtain a system of equations for $\psi_n(\xi)$:

$$\left[h_0(\xi)+V_{00}(\xi)\right]\psi_0(\xi)+\sum_{n}V_{0n}(\xi)\psi_n(\xi)=\eta\psi_0(\xi) \ , \tag{5a}$$

$$\left[h_0(\xi)+V_{nn}(\xi)\right]\psi_n(\xi)+\sum_{n'\neq n}V_{nn'}(\xi)\psi_{n'}(\xi)=\eta_n\psi_n(\xi)-V_{n0}(\xi)\psi_0(\xi) \ , \tag{5b}$$

where $n, n' \neq 0$ (also below), $\eta \equiv \eta_0 = E-\varepsilon_0$,

$$\eta_n \equiv E-\varepsilon_n \ , \quad \varepsilon_n \equiv \sum_{k}\varepsilon_{n_k} \ , \quad V_{nn'}(\xi)=\sum_{k}\left[V_{k0}^{nn'}(\xi)+\sum_{l>k}V_{kl}^{nn'}\right] , \tag{6}$$

$$V_{k0}^{nn'}(\xi)=\int_{\Omega_Q}dQ\Phi_n^*(Q)V_{k0}(q_k,\xi)\Phi_{n'}(Q), \quad V_{kl}^{nn'}=\int_{\Omega_Q}dQ\Phi_n^*(Q)V_{kl}(q_k,q_l)\Phi_{n'}(Q), \tag{7}$$

and we have separated the equation for $\psi_0(\xi)$ describing the generalised ground state of system elements, i. e. system state with minimum energy and complexity.

If we *avoid any perturbative reduction* of a problem and try to solve eqs. (5) by expressing $\psi_n(\xi)$ through $\psi_0(\xi)$ from eqs. (5b) with the help of Green function and substituting the result into eq. (5a), we get the *effective existence equation* for $\psi_0(\xi)$ [1-12]:

$$h_0(\xi)\psi_0(\xi)+V_{eff}(\xi;\eta)\psi_0(\xi)=\eta\psi_0(\xi) \ , \tag{8}$$

where the *effective (interaction) potential (EP)*, $V_{eff}(\xi;\eta)$, is given by

$$V_{eff}(\xi;\eta)=V_{00}(\xi)+\hat{V}(\xi;\eta) \ , \quad \hat{V}(\xi;\eta)\psi_0(\xi)=\int_{\Omega_\xi}d\xi' V(\xi,\xi';\eta)\psi_0(\xi') \ , \tag{9a}$$

$$V(\xi,\xi';\eta)=\sum_{n,i}\frac{V_{0n}(\xi)\psi_{ni}^0(\xi)V_{n0}(\xi')\psi_{ni}^{0*}(\xi')}{\eta-\eta_{ni}^0-\varepsilon_{n0}} \ , \quad \varepsilon_{n0}=\varepsilon_n-\varepsilon_0 \ , \tag{9b}$$

and $\{\psi_{ni}^0(\xi)\}$, $\{\eta_{ni}^0\}$ are complete sets of eigenfunctions and eigenvalues respectively for a truncated system of equations originating from "homogeneous" parts of eqs. (5b):

$$\left[h_0(\xi)+V_{nn}(\xi)\right]\psi_n(\xi)+\sum_{n'\neq n}V_{nn'}(\xi)\psi_{n'}(\xi)=\eta_n\psi_n(\xi) \ . \tag{10}$$

The eigenfunctions $\{\psi_{0i}(\xi)\}$ and eigenvalues $\{\eta_i\}$ found from eq. (8) are used to obtain other state-function components and the general solution (see eq. (4)):

$$\Psi(\xi,Q)=\sum_{i}c_i\left[\Phi_0(Q)\psi_{0i}(\xi)+\sum_{n}\Phi_n(Q)\psi_{ni}(\xi)\right] , \tag{11}$$

$$\psi_{ni}(\xi)=\int_{\Omega_\xi}d\xi' g_{ni}(\xi,\xi')\psi_{0i}(\xi'), \quad g_{ni}(\xi,\xi')=V_{n0}(\xi')\sum_{i'}\frac{\psi_{ni'}^0(\xi)\psi_{ni'}^{0*}(\xi')}{\eta_i-\eta_{ni'}^0-\varepsilon_{n0}} \ , \tag{12}$$

where coefficients $c_i$ should be found by state-function matching at the boundary where effective interaction vanishes. The observed system density, $\rho(\xi,Q)$, is obtained as state-function squared modulus, $\rho(\xi,Q)=|\Psi(\xi,Q)|^2$ (for "wave-like" complexity levels), or as state-function itself, $\rho(\xi,Q)=\Psi(\xi,Q)$ (for "particle-like" levels) [1].

Although the effective problem formulation of eqs. (8)-(12) is formally equivalent to initial equations (1), (2), or (5), it reveals in explicit form a *new quality* of *unreduced* problem solution, the *dynamic multivaluedness*, or *redundance*, *and entanglement* phenomenon. It emerges as an *excessive* number of locally *complete* and therefore *mutually incompatible* problem solutions, or system *realisations*, which, being *equally real*, are forced by the *same* driving interaction to *permanently replace* one another in a *dynamically random* order thus defined. Solution redundance follows e.g. from an elementary estimate of their number by the maximum power of characteristic equation for eigenvalues $\eta$ of the unreduced effective equation (8)-(9) [1-14]. It reflects a physically transparent phenomenon of *dynamic, or essential, nonlinearity* due to *dynamic links* of real interaction process expressed by the unreduced EP dependence on eigen-solutions to be found (eqs. (9)) and remaining "hidden" in usual problem formulation, eqs. (1)-(2), (5). Indeed, it is evident that unreduced interaction between, for example, two objects with *N* elements (eigenmodes) each will give at least $N^2$ element combinations ($N!$ combinations for *N* one-element objects), whereas there is only the same number of *N* places in reality for interaction products, which gives at least *N*-fold redundance of interaction products properly reflected by the above estimate.

By contrast, any usual, perturbative reduction of a problem towards its "exact", closed solution is equivalent to incorrect rejection of *all but one* system realisations (cf. standard "solution uniqueness" theorems). Such *dynamically single-valued*, or *unitary*, approximation of usual theory (including its versions of complexity, chaoticity, self-organisation, etc.) is equivalent to a *zero-dimensional, point-like projection* of dynamically multivalued, "multi-dimensional" reality, which explains a strange combination of usual science successes and its "unsolvable", stagnating old and growing new problems [1-10].

Thus discovered *intrinsic*, genuine *chaoticity* of *unceasing* realisation change is expressed by the really complete, *dynamically probabilistic* problem solution providing system density $\rho(\xi,Q)$ as a *causally probabilistic sum* of individual realisation densities, $\{\rho_{\mathrm{r}}(\xi,Q)\}$:

$$\rho(\xi,Q)=\sum_{\mathrm{r}=1}^{N_{\Re}}{}^{\oplus}\rho_{\mathrm{r}}(\xi,Q)\ . \tag{13}$$

where $N_{\Re}$ $(=N)$ is realisation number, while the $\oplus$ sign stands for the causally probabilistic sum meaning defined above. One obtains also the *dynamically derived*, *a priori* values of *realisation probabilities*, $\{\alpha_{\mathrm{r}}\}$:

$$\alpha_{\mathrm{r}}=\frac{N_{\mathrm{r}}}{N_{\Re}}\quad\left(N_{\mathrm{r}}=1,...,N_{\Re};\sum_{\mathrm{r}}N_{\mathrm{r}}=N_{\Re}\right),\quad\sum_{\mathrm{r}}\alpha_{\mathrm{r}}=1\ , \tag{14}$$

where $N_{\mathrm{r}}$ is the number of elementary realisations remaining often unresolved in the r-th "compound" realisation. It is important that eqs. (13), (14) contain not only the standard "expectation" value for large event series, but remain valid for any *single* event occurrence and even before it, providing *a priori probability* and its *universal dynamic origin*.

A related specific property of the obtained *unreduced* problem solution (and *real* system structure) is *dynamic entanglement* of interacting components appearing as a sum of dynamically weighted functions of interacting degrees of freedom $(\xi,Q)$ (see eqs. (11)-(12)). Being further amplified by a *probabilistically fractal* realisation structure (see below), it gives rise to a mathematically *rigorous* definition of tangible *quality* of *any* real structure that does *not* exist in the standard, unitary theory necessarily operating with merely *abstract*, "bodiless" imitations of reality within its *separable*, exact-solution "model". Indeed, system components can be physically disentangled, separated within any "closed" solution of usual theory, whereas such separation is impossible for a real, dynamically entangled structure. Those two major properties of unreduced problem solution (and real system structure it describes), redundant realisation number and dynamic entanglement within each realisation structure, are inseparably related to each other, so that any real interaction result (and meaning) can be described as *dynamically multivalued entanglement* of interacting entities.

*Dynamic complexity*, *C*, can be *universally* defined now as a growing function of system realisation number, or rate of change, equal to zero for only one realisation: $C = C(N_{\Re})$, $dC/dN_{\Re} > 0$, $C(1) = 0$. It is the latter, unrealistic case of zero unreduced complexity that is invariably considered in canonical, dynamically single-valued theory (*including* its *imitations* of complexity), which explains its old and new difficulties at various levels of reality [1]. Unreduced dynamic complexity is expressed by the majority of actually measured quantities (now *properly understood* as unreduced complexity measures), such as energy (temporal rate of realisation change in a *chaotic* order), mass (proportional to energy), momentum (spatial rate of realisation change), various charges, action, and entropy, now provided with a *unified* and *essentially nonlinear* interpretation in terms of underlying *interaction* [1-10] (see also below). Note that the universal complexity definition thus *derived* is based essentially on the obtained *causally complete, dynamically probabilistic solution* of *unreduced interaction problem*, eqs. (8)-(14), that *cannot* be replaced by formal counting of arbitrary, empirically inserted entities (as it is done in unitary science). It explains, in particular, why the unreduced complexity represents also a universal measure of *genuine* and omnipresent *chaoticity*.

The *whole* diversity of world structures is reproduced by a unified classification of realisation change regimes [1-10]. As can be seen from the unreduced EP formalism, eqs. (8)-(12), *similar* parameters of interacting entities (conveniently represented by characteristic system frequencies) lead to *essentially different* realisations and their relatively frequent change ($N_{\rm r} \approx 1$ and $\alpha_{\rm r} \approx 1/N_{\Re}$ for all r in eq. (14)), giving rise to strong, *global*, or *uniform chao*s regime. The opposite case of *sufficiently different* characteristic parameters of interaction process is reduced to *dynamically multivalued self-organisation* including extended, *chaotic* version of *self-organised criticality (SOC)* and *chaotic* change of *similar* realisations giving rise to more distinct structures and (externally) regular dynamics. The *whole* diversity of existing, *always* complex (dynamically multivalued) structures and motions is obtained by system parameter variation between those two limiting cases (including situations with several or many complexity levels, see below). Specifically, the transition to strong, global chaos regime occurs when the *chaoticity parameter*, $\kappa$, is close to unity:

$$\kappa \equiv \frac{\Delta\eta_{\rm i}}{\Delta\eta_{\rm n}} = \frac{\omega_\xi}{\omega_Q} \simeq 1 \ , \qquad (15)$$

where $\Delta\eta_{\rm i}, \omega_\xi$ and $\Delta\eta_{\rm n} \sim \Delta\varepsilon, \omega_Q$ are characteristic energy-level separations and frequencies for inter-component and intra-component system motions, respectively. At $\kappa \ll 1$ one obtains a relatively regular structure/motion of *multivalued* SOC regime, while at $\kappa$ growing from 0 to 1 one obtains a series of transitions to progressively more chaotic regimes culminating at $\kappa \simeq 1$ in a global chaos regime (at $\kappa > 1$ one gets transitions to ever more regular regimes but with a different, "inverse" system configuration) [1-5]. Besides the evident advantages of obtained *universal* classification of *any* dynamics, one may note such extensions with respect to usual, unitary picture as existence of *permanent* and *chaotic* realisation change within *any*, even externally "regular" structure or motion (providing the *dynamic origin of time* and truly *universal entropy growth principle*, see below) and unification within our chaotic SOC regime of extended (multivalued and chaotic) versions of various separated cases of usual "science of complexity", such as self-organisation, SOC, adaptability, various "synchronisation" cases, mode locking, and fractality.

The unreduced, *dynamically probabilistic fractality* represents *essential*, complex-dynamic extension of usual fractal hierarchy and is obtained as inevitable development and *ultimately complete* content of dynamic entanglement (*nonseparability*) and complexity [1,5,7,8]. It appears due to EP dependence on *unknown* solutions of truncated system of equations, eqs. (10), expressing problem *nonintegrability*. The generalised EP formalism of eqs. (8)-(9) can be applied now to the truncated system (10), transforming it into a single effective equation:

$$\left[h_0(\xi)+V_{\mathrm{eff}}^{\mathrm{n}}(\xi;\eta_{\mathrm{n}})\right]\psi_{\mathrm{n}}(\xi)=\eta_{\mathrm{n}}\psi_{\mathrm{n}}(\xi)\ , \tag{16}$$

where the second-level EP action is similar to the combined version of eqs. (9),

$$V_{\mathrm{eff}}^{\mathrm{n}}(\xi;\eta_{\mathrm{n}})\psi_{\mathrm{n}}(\xi)=V_{\mathrm{nn}}(\xi)\psi_{\mathrm{n}}(\xi)+\sum_{\mathrm{n}'\neq\mathrm{n},\ \mathrm{i}}\frac{V_{\mathrm{nn}'}(\xi)\psi_{\mathrm{n}'\mathrm{i}}^{0\mathrm{n}}(\xi)\int\limits_{\Omega_\xi}d\xi'\psi_{\mathrm{n}'\mathrm{i}}^{0\mathrm{n}*}(\xi')V_{\mathrm{n}'\mathrm{n}}(\xi')\psi_{\mathrm{n}}(\xi')}{\eta_{\mathrm{n}}-\eta_{\mathrm{n}'\mathrm{i}}^{0\mathrm{n}}+\varepsilon_{\mathrm{n}0}-\varepsilon_{\mathrm{n}'0}}\ , \tag{17}$$

and $\{\psi_{\mathrm{n}'\mathrm{i}}^{0\mathrm{n}}(\xi),\eta_{\mathrm{n}'\mathrm{i}}^{0\mathrm{n}}\}$ is the eigen-solution set for a second-level truncated system:

$$h_0(\xi)\psi_{\mathrm{n}'}(\xi)+\sum_{\mathrm{n}''\neq\mathrm{n}'}V_{\mathrm{n}'\mathrm{n}''}(\xi)\psi_{\mathrm{n}''}(\xi)=\eta_{\mathrm{n}'}\psi_{\mathrm{n}'}(\xi)\ ,\ \ \mathrm{n}'\neq\mathrm{n}\ ,\ \ \mathrm{n},\mathrm{n}'\neq 0\ . \tag{18}$$

Similar to the first-level EP of eqs. (8)-(9), the *essentially nonlinear* EP dependence on the eigen-solutions to be found in eqs. (16)-(17) leads to the second-level splitting (now of truncated system solutions) into many incompatible realisations (numbered by index $\mathrm{r}'$):

$$\left\{\psi_{\mathrm{ni}}^{0}(\xi),\eta_{\mathrm{ni}}^{0}\right\}\to\left\{\psi_{\mathrm{ni}}^{0\mathrm{r}'}(\xi),\eta_{\mathrm{ni}}^{0\mathrm{r}'}\right\}\ . \tag{19}$$

That hierarchy of multivalued realisation levels continues until we finally get one integrable equation for one function "closing" problem solution. The obtained *truly complete* solution is represented by the *dynamically probabilistic fractal* providing the ultimately exact, *dynamically unified* structure of real-world complexity, at any its level or part:

$$\rho(\xi,Q)=\sum_{\mathrm{r},\mathrm{r}',\mathrm{r}''...}^{N_{\Re}}{}^{\oplus}\rho_{\mathrm{rr}'\mathrm{r}''...}(\xi,Q)\ , \tag{20}$$

where indexes $\mathrm{r},\mathrm{r}',\mathrm{r}'',...$ enumerate chaotically changing realisations of consecutive levels of *dynamic (probabilistic) fractality*. The time-averaged *expectation value* for the dynamically fractal system density is obtained as

$$\rho(\xi,Q)=\sum_{\mathrm{r},\mathrm{r}',\mathrm{r}''...}^{N_{\Re}}\alpha_{\mathrm{rr}'\mathrm{r}''...}\rho_{\mathrm{rr}'\mathrm{r}''...}(\xi,Q)\ , \tag{21}$$

where *dynamically determined probabilities* reproduce eqs. (14) at various fractality levels:

$$\alpha_{\mathrm{rr'r''}\ldots} = \frac{N_{\mathrm{rr'r''}\ldots}}{N_{\Re}} \ , \qquad \sum_{\mathrm{r,r',r''}\ldots} \alpha_{\mathrm{rr'r''}\ldots} = 1 \ . \tag{22}$$

Note that contrary to any perturbation theory expansion, eqs. (20)-(22) (with the underlying EP formalism of eqs. (8)-(19)) provide the *truly exact*, totally realistic problem solution, where each term describes a really emerging structure element. The resulting dynamical fractal is an essential extension of canonical, purely abstract (mathematical) and dynamically single-valued fractals: the former does not possess, in general, the simplified scale symmetry of usual fractals and describes the property of *probabilistic dynamic adaptability* and *purposeful evolution* of *all* real structures due to permanent, *interaction-driven* realisation change. It reflects a deeper real-world symmetry, the *universal symmetry of complexity*, that actually determines any real (complex) system dynamics and evolution [1-5] (see below).

A major feature of unreduced complex-dynamic fractality (absent in its unitary imitations) is its very big, *exponentially huge operation power* (e.g. state change rate) due to *autonomous* creativity and realisation change driven only by the main interaction [5-10]. If $N = N_{\mathrm{unit}} n_{\mathrm{link}}$ is the total number of interaction links in the system (where $N_{\mathrm{unit}}$ is the number of interacting units and $n_{\mathrm{link}}$ the number of links per unit), then the system operation power *P*, proportional to it realisation number $N_{\Re}$, is determined by the *total number of link combinations*, i.e. $N_{\Re} \simeq N!$: $P \propto N_{\Re} \simeq N! \simeq \sqrt{2\pi N}(N/e)^N \sim N^N$. As *N* is a large number itself (for example, $N \geq 10^{12}$ for brain or genome interactions [5,7,8]), one obtains really huge, practically infinite *P* values. Any unitary (basically regular and sequential) model of the same system has the power $P_0$ that can grow only as $N^{\beta}$ ($\beta \sim 1$), so that $P/P_0 \sim N^{N-\beta} \sim N^N \to \infty$, which clearly demonstrates the advantages of complex-dynamic operation regime and the origin of "magic" properties of living and intelligent systems.

## SYMMETRY OF COMPLEXITY AND THE DYNAMIC ORIGIN OF TIME

The *emergent*, axiom-free structure of the universal science of complexity reflects the fact that any real entity is *explicitly obtained* as a result of interaction process development, in exact agreement with its real emergence. Chaotically changing realisations and their groups form the observed diversity of world structures and dynamics (see previous section). Its most fundamental levels, including elementary particles with all their intrinsic and dynamic properties, space and time appear from the simplest possible interaction configuration, the one of two attracting, initially homogeneous (and physically real) fields or media specified as electromagnetic and gravitational protofields [1,4,5,13,14]. Further structural levels progressively emerge through interaction of lower-level structures and form the hierarchy of *complexity levels* or probabilistic dynamical fractal of the *unified* world structure.

Each complexity level produces the corresponding level of *physically real, emergent space* and *naturally flowing time*. The *elementary space distance*, $\Delta x$, is *explicitly obtained* as eigenvalue separation found from eq. (8), $\Delta x = \Delta \eta_{\mathrm{i}}^{\mathrm{r}}$, where eigenvalue separation between neighbouring realisations gives the *elementary length*, or size of system jump between realisations, $\lambda = \Delta x_{\mathrm{r}} = \Delta_{\mathrm{r}} \eta_{\mathrm{i}}^{\mathrm{r}}$, while eigenvalue separation within one realisation determines the minimum *size of emerging structure* itself, $r_0 = \Delta x_{\mathrm{i}} = \Delta_{\mathrm{i}} \eta_{\mathrm{i}}^{\mathrm{r}}$. Thus, at the lowest complexity level $\lambda = \lambda_{\mathrm{C}}$ is the Compton wavelength determining internal complex dynamics, mass and electromagnetic interactions of the electron, while $r_0$ is the "classical radius of the electron" provided now with a well-specified physical meaning (the heaviest particles give extreme values of space elements coinciding in our theory with the consistently modified Planckian length unit) [4,13]. Note that any level of space structure has a *causally quantised*, *dynamically discrete* texture (as opposed to *both* usual space continuity *and* its arbitrary, postulated discreteness). The *tangible*,

"material" nature of space is due to (fractal) *dynamic entanglement* of interacting entities (see above) represented by the two protofields at the lowest level of "embedding" physical space.

The *elementary time interval*, $\Delta t$, is explicitly obtained through *intensity*, in the form of *frequency*, $\nu$, of causally specified *events* of *realisation emergence/change* (the latter directly following from *causally derived*, *universal* dynamic multivaluedness): $\Delta t = \tau = 1/\nu$. Time leap $\Delta t = \tau$ is none other than duration of system jump across the elementary length $\lambda = \Delta x_{\rm r}$ and therefore can be expressed as $\tau = \lambda/\upsilon_0$, where $\upsilon_0$ is the material signal propagation speed in a lower-level component structure. It is important that we obtain thus a physically real, *permanently flowing* and *dynamically irreversible*, but *not* materially tangible time (it *cannot* therefore be really "mixed" with space in a unified "manifold", contrary to formal time of unitary theory). Unceasing time flow is due to *incompatible realisation change* driven by the underlying interaction itself, while impossibility of its mechanical reversion follows from *dynamic randomness* of realisation change, which demonstrates a deep, intrinsic *connection between time and randomness* completely ignored by unitary science paradigm that tends instead to consider time as a manifestation of basic *regularity* of world dynamics (and remains thus with a stagnating *mystery* of irreversible time flow).

As emerging, physically real space and time thus obtained contain major material and dynamic properties of appearing complex-dynamic structure, a natural and *universal complexity measure* is provided by the simplest function proportional independently to space and time intervals, i.e. (generalised) *action* $\mathcal{A}$ [1-10,13,14]. Intrinsic discreteness of complex dynamics (system jumps between incompatible realisations) gives rise to *causal quantisation* of action:

$$\Delta\mathcal{A} = p\Delta x - E\Delta t \quad , \tag{23}$$

where *p* and *E* are initially just coefficients, but immediately recognised, by analogy to classical mechanics, as generalised *momentum* and (total) *energy*:

$$p = \frac{\Delta\mathcal{A}}{\Delta x}\Big|_{t=\text{const}} = \frac{\mathcal{A}_0}{\lambda}, \tag{24}$$

$$E = -\frac{\Delta\mathcal{A}}{\Delta t}\Big|_{x=\text{const}} = \frac{\mathcal{A}_0}{\tau}, \tag{25}$$

where $\mathcal{A}_0$ is a characteristic action value (by modulus) at a given complexity level. It follows that action is *integral*, while momentum and energy are *differential complexity measures*. Note that here and below *p*, *x* and other quantities can be considered as vectors if necessary.

Irreversible time flow ($\Delta t > 0$) and positive total energy/mass ($E > 0$) in eq. (25) imply that $\Delta\mathcal{A} < 0$, i.e. generalised action always *decreases* in real interaction process development. It represents therefore a *potential*, *latent* form of dynamic complexity called *dynamic information* $I = \mathcal{A}$ and given to the system at birth, *before* actual interaction development. The appearing new structure is characterised by a complementary form of complexity, *dynamic entropy S*, which describes the *implemented*, *unfolded* structure complexity. A permanent aspect of a system (interaction process) is provided by its *total complexity* $C = \mathcal{A} + S$ that remains thus *unchanged* in any system evolution:

$$C = \mathcal{A} + S = \text{const}, \quad \Delta S = -\Delta\mathcal{A} > 0 \quad . \tag{26}$$

The obtained absolutely *universal law of conservation, or symmetry, of complexity* naturally includes, therefore, *both conservation* of total complexity *and unceasing change* of its form, from *decreasing* dynamic information/action to *increasing* dynamic entropy (= generalisation of entropy growth principle) [1,3-5]. It means that due to dynamic multivaluedness *any*, even externally "regular" structure emergence reflects definite entropy *growth*, as opposed to the respective contradictory situation in usual, dynamically single-valued theory.

Complexity conservation law can be provided with a unified differential expression, the *generalised Hamilton-Jacobi equation*, if we divide the second part of eq. (26) by $\Delta t\big|_{x=\text{const}}$ :

$$\frac{\Delta \mathcal{A}}{\Delta t}\Big|_{x=\text{const}} + H\left(x, \frac{\Delta \mathcal{A}}{\Delta x}\Big|_{t=\text{const}}, t\right) = 0 \ , \quad H > 0 \ , \tag{27}$$

where the Hamiltonian, $H = H(x,p,t)$, is a differential measure of unfolded, entropic form of complexity, $H = (\Delta S/\Delta t)\big|_{x=\text{const}}$, and the condition of its positivity ($H > 0$) provides a universal expression of the *arrow/direction of time* (towards permanently growing dynamic complexity-entropy) [4,5,13,14]. Since action increment is proportional to that of *generalised wavefunction, or distribution function* $\Psi(x,t)$, $\Delta \mathcal{A} = -\mathcal{A}_0 \Delta \Psi/\Psi$ (*causal quantisation*) [1,3-5,8-10,13,14], eq. (27) has its dual partner, *generalised Schrödinger equation* for $\Psi(x,t)$:

$$\mathcal{A}_0 \frac{\Delta \Psi}{\Delta t}\Big|_{x=\text{const}} = \hat{H}\left(x, \frac{\Delta}{\Delta x}\Big|_{t=\text{const}}, t\right)\Psi(x,t) \ , \tag{28}$$

where the Hamiltonian operator $\hat{H}$ is obtained from its functional form $H$ with help of causal quantisation. The generalised wavefunction describes a specific, *transiently disentangled* system state during its transitions between regular, entangled realisations and can be obtained as a part of unreduced general solution of a problem [1]. At the lowest, *quantum* levels of world complexity it provides a *causally complete*, *realistic* extension of usual quantum-mechanical wavefunction, now liberated from its postulated "mysteries" [1,4,5,13,14]. The above causal quantisation rule describes the *real* process of system *transition* between realisations giving rise to causally extended, universal version of another "quantum" postulate, *Born's probability rule*, $\alpha_\text{r} = |\Psi(X_\text{r})|^2$ (with the r-th realisation configuration $X_\text{r}$), that provides another expression for realisation probabilities $\alpha_\text{r}$ (cf. eqs. (14)). At higher complexity levels $\Psi(x,t)$ takes the form of distribution function determining realisation probabilities directly ("particle-like" complexity levels) or by its squared modulus ("wave-like" levels).

Expanding the Hamiltonian $H = H(x,p,t)$ in power series on momentum $p$ and using the result in eqs. (27), (28), it is easy to see [1,4,5] that these equations represent a unified extension of *all* usual, "model" (both "linear" and "nonlinear") equations of unitary theory, confirming universality of the obtained *Hamilton-Schrödinger formalism* and the underlying *symmetry of complexity*. The latter actually describes the symmetry between all progressively taken system realisations at a given complexity level (the symmetry of complexity constitutes thus the *basis of system dynamics* as such), while general, multi-level complexity unfolding provides the *universal evolution law* and *direction/meaning of progress* [3,4]. It is important that contrary to regular but always finally "broken" (inexact) symmetries *formally postulated* in unitary science, the universal symmetry of complexity is always *exact* (unbroken) but *irregularly* structured symmetry *equivalent to system dynamics and evolution*. Therefore it gives rise to essential extension of exact-science methods and a real possibility to obtain *rigorous* and totally *consistent (complete)* description of higher-complexity systems and phenomena [1] usually only studied by purely empirical methods in humanities.

## UNREDUCED MATHEMATICS OF REAL WORLD DYNAMICS

We have already noted above that the intrinsically realistic, unreduced basis of the universal science of complexity allows for its direct expression in intuitively transparent, generally understandable form (as opposed to irreducible abstraction and mystification of unitary knowledge projection). On the other hand, that realism is obtained due to the *truly rigorous*, non-simplified mathematical description that does reveal *qualitatively new* phenomenon of *dynamically multivalued entanglement* of any real interaction components, giving rise to universal dynamic complexity and its various manifestations. In this section we provide a summary of distinctive features of that *unreduced mathematics of complexity* obtained simply due to the truly consistent analysis of arbitrary interaction process avoiding any "model" approximations of usual mathematical framework (including its approaches to complexity).

Note, first of all, the explicitly, *intrinsically unified mathematical basis* of the universal science of complexity including the *universal symmetry of complexity* as the *single*, unified and always valid *law* that gives rise to all particular laws, equations, etc. and the corresponding *single mathematical structure* of *dynamically probabilistic fractal* obtained as the *causally complete solution* of unreduced interaction problem (see previous sections). Both these aspects of obtained intrinsically unified knowledge are in agreement with the physically unified nature of observed reality and in strong contrast to irreducible ruptures of unitary, projective *imitations of reality* (whose true origin is now clearly specified), including recent attempts of their artificial, inconsistent joining that *cannot*, contrary to *intrinsic* unity, provide a *real* problem solution.

We emphasise the following *emergent*, *rigorously derived* features of the unreduced mathematics of real, complex world dynamics (see eqs. (8)-(22) and respective explanations):

(1) *Nonuniqueness* of any real problem solution is expressed by the major property of *dynamic multivaluedness*. It is a qualitative extension of conventional solution uniqueness, the latter being obtained due to a characteristic "vicious circle" trap of reductive, perturbative approach (where *assumed* uniqueness of interaction potential leads to uniqueness of solution, while that "natural" initial assumption appears eventually to be *wrong* with respect to real, effective interaction potential). Note that dynamic multivaluedness should be clearly distinguished from its various unitary imitations, such as semi-empirical "multistability" or purely abstract "strange attractors", which are obtained by an artificially "intricate" structure of usual, dynamically single-valued solution (similar to a strongly entangled one-dimensional thread) and describe *compatible*, coexisting parts of *one and the same* (reduced) solution.

(2) Clear mathematical expression of explicit *dynamic emergence* of structures, well-specified *dynamic origin* of *events* and *time* are obtained in the form of *unceasing* realisation change and can be expressed by the general property of *absence of self-identity* of *any* mathematical (as well as real) structure, $\mathfrak{A} \neq \mathfrak{A}$, as opposed to absolute domination of (often implicit) self-identity postulate in usual mathematical constructions.

(3) Fractally structured (and chaotically changing) *dynamic entanglement* of interacting entities provides a *rigorous* expression of tangible *quality* of emerging structures (again totally absent in unitary mathematics dealing exclusively with "immaterial", abstract imitations of reality).

(4) Absence of "exact", closed solutions to real interaction (or many-body) problems has a clearly specified origin, dynamic multivaluedness (see item (1)), providing a *unified*, *causally complete* meaning of *randomness*, *nonintegrability*, *nonseparability*, *noncomputability*, etc.

(5) *Dynamic discreteness*, or *causal quantisation*, of real interaction dynamics follows directly from its *unreduced* development and involves solution (and thus system evolution) *nonunitarity* and causal, dynamically discrete *origin of space*. Dynamic discreteness should be distinguished from any kind of formal, artificially imposed discreteness or postulated quantisation.

Note that the *intrinsically unified* nature of the extended mathematics of complexity, its *exact* correspondence to *real* world structure/dynamics and all its particular aspects (1)-(5) express the *inbred interdisciplinarity* of the universal science of complexity and reveal the true, equally deep origin of *disciplinary*, irreducibly broken structure of traditional, unitary knowledge.

## PROBLEM-SOLVING APPLICATIONS OF THE UNIVERSAL SCIENCE OF COMPLEXITY

All the above properties and advantages of the unreduced problem solution obtain their definite confirmation by successful application of the proposed framework to various particular, otherwise stagnating problems at different levels of complexity, from fundamental physics (elementary particles and cosmology) to emerging consciousness and its most sophisticated products usually studied in humanities [1-15]. It is important to emphasise that we are talking here not about usual reduced imitations (or "modelling") of real phenomena, but about their exact, causally complete description explicitly solving "unsolvable" problems.

Before proceeding with a brief summary of particular application result, let us specify a few general *principles of complex system operation and design* emerging in all applications [10]:

(1) *Complexity correspondence principle* refers to efficient interaction between units of comparable complexity [1,4,5]. In particular, higher complexity enslaves lower complexity, while low-complexity tools *cannot* control/simulate higher complexity behaviour.

(2) *Complex-dynamic control principle* emphasises *complexity development* (from dynamic information to entropy) as the purpose of efficient, *creative* control (as opposed to usual, restrictive control paradigm), leading to the universal concept of *unreduced sustainability* [9].

(3) *Unreduced interaction principle* reveals the huge power of natural interaction complexity, in the form of *exponentially huge operation power* of dynamically probabilistic (chaotic) fractal underlying the "mysteries" of life, intelligence and consciousness [4-10].

The ***first group of applications*** includes *unified complex-dynamic, causal origin of the Universe, its laws, fundamental structures and their properties* [1,4,5,13,14]. Space, time, elementary particles and fields, their properties are explicitly derived from unreduced solution of interaction problem with the simplest possible configuration (two attracting, initially homogeneous protofields). One gets totally realistic (causally complete) and intrinsically unified versions of quantum mechanics and relativity, with all usual, postulated "mysteries" explained as being due to artificial reduction of standard unitary approach. New and old "difficult" problems of standard theory (dark matter/energy, particle mass spectrum) have the same origin and are solved now (or even do not appear), without artificial introduction of "hidden" and abstract entities (dimensions, particles, fields, etc.). As a result, the real world structure is consistently obtained as it is, not as an abstract "model".

The ***second group of applications*** provides a consistent description of *irreducibly complex dynamics of real nanobiosystems* [5,6]. We show that genuine, dynamically multivalued and strong chaoticity is *inevitable* at the smallest scales of atomic and molecular interactions and consistently explains the "magic" properties of biosystems. We explicitly demonstrate the fundamental, purely dynamic origin of *genuine quantum chaos* [1,5,6,11,12], thus solving the whole bundle of related problems. *Complex dynamics of causal quantum measurement* [1] and *dynamic emergence of classical behaviour* in simplest bound systems (multivalued SOC regime) [1,13] complete the *consistent, universal basis for nanobioscience applications*.

The ***third group of applications*** further extends these results to foundation of *complex-dynamic biology* emerging now as an *exact, causally complete* kind of science, with its major applications to *reliable genetics* and *integral medicine* [1,5,7]. Our dynamically probabilistic fractal, obtained as *unreduced solution* to many-body interaction problem, shows major observed properties of a *living* structure, including *autonomic adaptability*, *purposeful dynamics*, objectively specified *birth, life, and death of a system*, and *exponentially huge operation power*. *Reliable genetics* is based on the *unreduced* analysis of all relevant *genome interactions*, and we show rigorously that their practically *total neglect* in the dominating empirical approach inevitably creates a delayed-time genetic bomb with catastrophic consequences. *Integral medicine* extends our approach to the whole system of each *individual* organism interactions and includes complex-dynamic, reliable and creative control (see the above principle (2)) of the resulting dynamically probabilistic fractal of life.

The ***fourth group of applications*** provides the causally complete understanding of *emerging genuine intelligence and consciousness*, in both natural and artificial systems [1,8]. We show that unreduced intelligence and consciousness are natural properties of *high enough*, well-specified levels of *unreduced* interaction complexity, absent in any its unitary imitation by "involved" but regular dynamics. We obtain thus the unique way of creation, principles of design and specific properties of *artificial* but *genuine* intelligence and consciousness, including a major property of exponentially huge power (complex-dynamic parallelism).

The ***fifth group of applications*** deals with creation of new, *complex-dynamic information systems*, including *intelligent software* and *autonomic communication networks* [10]. The related *complexity transition* in operation and design of information and communication systems increases dramatically their operation power towards the new quality of "living" and intelligent behaviour, including such applications as *knowledge-based networks*.

The ***sixth group of applications*** deals with the necessary *transition to genuine sustainability* at a *superior level of civilisation complexity*, or *revolution of complexity* [1,9]. We *rigorously* substantiate the necessity of transition as *unified* and *unique* issue from the current ecological and development crisis and consider various aspects of the proposed solution, such as the underlying *new kind of knowledge* of the universal science of complexity with its rigorously specified *concept of progress*, *complexity-increasing production*, *non-unitary social structure*, and *new kind of settlement*.

The ***seventh group of applications*** deals with consistent description of superior complexity levels usually studied in the humanities, such as ethics, aesthetics, and spiritual issues [1]. We provide *rigorously specified* concepts of *good* (as optimally growing complexity-entropy), *beauty* (as relative attained complexity), and *spirit* (as unreduced superior-level complexity).

One should emphasise the important unifying role of the *new content and organisation of science* oriented to *unreduced* understanding of *complex-dynamic* reality within a *liberal, creative structure* of knowledge production and exchange [1,15]. One obtains thus a holistic, complex-dynamic universe picture, from the lowest to highest complexity levels, proving its efficiency by *explicit solution* of otherwise *stagnating* problems at all levels. In that way, the universal science of complexity demonstrates the *provably unique* (and efficient) opportunity of genuine knowledge progress towards its intrinsically unified, creative and ultimately useful state, giving a well-specified positive answer to our initial inquiry (see Introduction).

## ACKNOWLEDGMENTS

The author is greatful to organisers of DECOS-2006 conference for invitation and financial support.